\documentclass[conference]{IEEEtran}

\usepackage{amssymb,amsmath,amsfonts}
\usepackage[]{algorithm2e}
\usepackage{blkarray}
\usepackage{bbm}
\usepackage{bm}
\usepackage{cases}
\usepackage{cite}
\usepackage[usenames,dvipsnames]{color}
\usepackage{dsfont}
\usepackage{float}
\usepackage{flushend}
\usepackage{graphicx}
\usepackage{hyperref} 
\usepackage[latin9]{inputenc}
\usepackage{mathtools}
\usepackage{mathrsfs}
\usepackage{setspace}
\usepackage{soul} 
\usepackage{subfigure}
\usepackage{stfloats}
\usepackage{tikz}
\usepackage{url}
\usepackage{verbatim}
\usepackage{xspace}
\usepackage{algpseudocode}

\usepackage{caption,epstopdf}

\newtheorem{thm}{Theorem}

\def\mb{\mathbf}

\def\mc{\mathcal}

\makeatletter
 \let\old@ps@headings\ps@headings
 \let\old@ps@IEEEtitlepagestyle\ps@IEEEtitlepagestyle
 \def\confheader#1{%
 \def\ps@headings{%
 \old@ps@headings%
 \def\@oddhead{\strut\hfill#1\hfill\strut}%
 \def\@evenhead{\strut\hfill#1\hfill\strut}%
 }%
 \def\ps@IEEEtitlepagestyle{%
 \old@ps@IEEEtitlepagestyle%
 \def\@oddhead{\strut\hfill#1\hfill\strut}%
 \def\@evenhead{\strut\hfill#1\hfill\strut}%
 }%
\ps@headings%
 }
\makeatother
\confheader{
\small{Proceedings of the 10th RSI International Conference on Robotics and Mechatronics (ICRoM 2022), Nov. 15-18, 2022, Tehran, Iran} 
}
\usepackage[pscoord]{eso-pic}
\newcommand{\placetextbox}[3]{
\setbox0=\hbox{#3}
\AddToShipoutPictureFG*{ \put(\LenToUnit{#1\paperwidth},\LenToUnit{#2\paperheight}){\vtop{{\null}\makebox[0pt][c]{#3}}}
}
}
\placetextbox{.23}{0.055}{\textbf{\small{978-1-6654-5452-0/22/\$31.00~\copyright 2022 IEEE}}}

\begin{document}

\title{\bf Consensus-based Networked Tracking in Presence of Heterogeneous Time-Delays 
} 

\author{\IEEEauthorblockN{ Mohammadreza Doostmohammadian}
\IEEEauthorblockA{\textit{School of Electrical Engineering}, \\ \textit{Aalto University}, 
 Espoo, Finland, \\
\texttt{name.surname@aalto.fi},\\
\textit{Mechatronics Department}, \\ 
\textit{Faculty of Mechanical Engineering}, \\
\textit{Semnan University}, Iran, \\ \texttt{doost@semnan.ac.ir}}
\and
\IEEEauthorblockN{Mohammad Pirani}
\IEEEauthorblockA{\textit{Mechanical and Mechatronics Eng}, \\ \textit{University of Waterloo}, 
Canada, \\
\texttt{pirani@uwaterloo.ca}}
\and
\IEEEauthorblockN{Usman A. Khan}
\IEEEauthorblockA{\textit{ Department of Electrical  Eng}, \\ 
\textit{Tufts University}, Medford, MA, USA,\\
\texttt{khan@ece.tufts.edu}
}
}

\maketitle
\begin{abstract}
    We propose a distributed (single) target tracking scheme based on networked estimation and consensus algorithms over static sensor networks. The tracking part is based on \textit{linear} time-difference-of-arrival (TDOA) measurement proposed in our previous works. This paper, in particular, develops delay-tolerant distributed filtering solutions over sparse data-transmission networks. We assume general arbitrary heterogeneous delays at different links. This may occur in many realistic large-scale applications where the data-sharing between different nodes is subject to latency due to communication-resource constraints or large spatially distributed sensor networks. The solution we propose in this work shows improved performance (verified by both theory and simulations) in such scenarios. Another privilege of such distributed schemes is the possibility to add localized fault-detection and isolation (FDI) strategies along with survivable graph-theoretic design, which opens many follow-up venues to this research. To our best knowledge no such \textit{delay-tolerant} distributed linear algorithm is given in the existing distributed tracking literature.
\end{abstract}


\section{Introduction} \label{sec_intro}
\IEEEPARstart{L}{ocalized} and distributed algorithms are an emerging field of research with a bright future in IoT and cloud-based applications. These algorithms are mostly based on the well-known consensus strategy primarily introduced in \cite{Olfati-2003}.  This research area has recently gained attention in filtering and estimation over networks and graph signal processing with different applications from system monitoring \cite{ilic2010modeling} to target tracking and localization \cite{safavi2018distributed}. In this paper, we consider distributed TDOA-based tracking with linear measurements that is shown to outperform the nonlinear counterparts in distributed scenarios, see details in \cite{tase,vtc_tracking}. Motivated by this, we further extended those research papers to address possible time delays in the data transmission networks among sensors. This is more realistic in practical large-scale applications in harsh environments and under limited communication and data-sharing resources.

\subsection{Background}
 Different distributed estimation and filtering methods have been considered in both signal processing and control literature with some applications in collaborative mobile robotic networks. The existing literature can be divided into two main categories: (i) double time-scale methods with an inner consensus-loop between every two consecutive sampling time steps of the system dynamics \cite{olfati:cdc09,he2019secure,battilotti2021stability}, and (ii) single time-scale methods with the same time-scale of filtering and system dynamics \cite{deghat2019detection,mohammadi2015distributed,usman_cdc:10,wang2019distributed,mitra2021distributed,Ugrinovskii8742899}. The scenario (ii) typically assumes system dynamic stability or the so-called local observability in the neighborhood of every sensor, which requires many direct communications to ensure system observability at every sensor. This implies a high rate of data transmission which is very restrictive in large-scale applications in terms of both infrastructure cost and budget for communication devices. Most importantly, this may add latency in data exchange and leads to high traffic over the network. On the other hand, in double time-scale scenario (ii), even though by high \textit{rate} of communication and information-sharing the system eventually becomes observable over any sparse networks, this comes with the cost of fast communication and processing devices. This is because in these methods the number of consensus and communication iterations needs to be much more than the network diameter. In contrast to the existing methods, a single time-scale scenario with distributed observability assumption is proposed in our previous works  \cite{jstsp,jstsp14} which significantly reduces the networking and processing times. Recall that the distributed observability assumption is shown to be much less restrictive as it does not need local observability in the neighborhood of any sensor but only global observability (as in centralized filtering) plus some mild assumptions on the communication network. This further motivates the target tracking application in this work. However, what is missing from these works is to consider \textit{delay-tolerant} algorithms over such networked scenarios. This gap is addressed in the current paper.

\subsection{Main Contributions}
Distributed estimation under \textit{heterogeneous} time-delays is primarily addressed in our recent work \cite{est_delay}. In the current work, we extend the results to a single target-tracking application based on TDOA measurements. This distributed range-based tracking scenario, recently adopted by \cite{ennasr2016distributed,ennasr2020time}, considers a \textit{nonlinear} setup in general. This nonlinearity results in filtering inaccuracy in two aspects; (i) most existing ``distributed'' estimation and observer setups are linear, which asks for the \textit{linearization} of this model at every iteration. (ii) The  \textit{linearized measurement matrix is time-varying and is a function of the position of the target} (which is not precisely known in general and needs to be estimated). Due to these drawbacks, the \textit{linearized} distributed tracking is not very accurate. What we suggested \cite{vtc_tracking,tase}, is a target-independent\footnote{Note that the TDOA measurements certainly depend on the target location, but the measurement ``matrix'' in our model is independent of that information.} and time-invariant linear measurement matrix (as in general LTI setups) which is compatible with most existing distributed filtering schemes. What is new in this paper is to take into account possible time delays in information exchange between every two sensors which makes the networked filtering scenario even more challenging. The delays, in general, are assumed to be arbitrary and heterogeneous (different) at different transmission links. As a typical assumption in literature, we consider a known upper-bound on the delays which implies no packet drops and information loss over the network. We prove error stability under such latency in network models and we prove and show by simulation that the estimation error decays sufficiently fast at all sensors (despite the time-delayed data-sharing) such that the proposed strategy acceptably tracks a single maneuvering target.  

\subsection{Paper Organization}
Section \ref{sec_setup} gives the general setup and preliminaries for the problem formulation. Section \ref{sec_main} provides the main results on the distributed tracking and delay-tolerance.
Section \ref{sec_simi} provides some illustrative simulations to show the performance of the proposed distributed filtering. Section \ref{sec_conc} concludes the paper. 

\subsection{General Notations}
 As a general notation, bold small letters denote the column vectors, small letters denote scalar variables, and capital letters present the matrices. $\mb{I}_N$  and $\mb{0}_N$ denote the identity and all-$0$s matrices of size $N$. $\mb{1}$ denotes the column vector of all-$1$s. ``;'' denotes the vector column concatenation. $\|\cdot\|$ denotes the 2-norm of a vector.  

\section{The Problem Setup} \label{sec_setup}
\begin{figure} [t]
		\centering
		\includegraphics[width=3.2in]{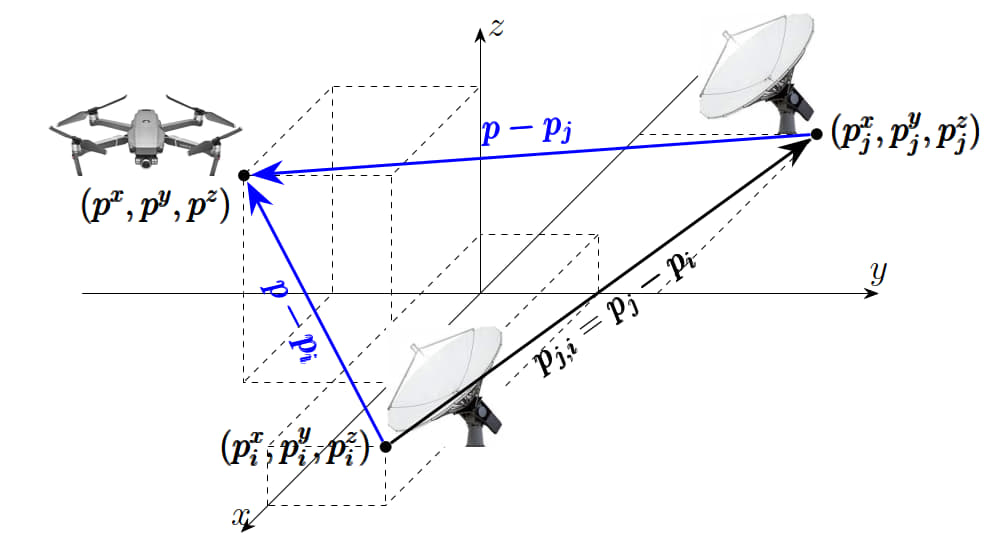}
		\caption{A group of static sensors (radars) tracking a mobile target (drone). 
		}
		\label{fig_radar_drone}
\end{figure}

\subsection{The Target Dynamics Model}
Our general tracking framework is shown in Fig.~\ref{fig_radar_drone}. As in many tracking literature, the target dynamics is assumed unknown and thus the following nearly-constant-velocity (NCV) dynamics\footnote{This work is not restricted to a particular model dynamics, but any other target dynamics model, for example, the NCA (nearly-constant-acceleration)  or the Singer model, can be applied. See more examples in \cite{bar2004estimation}.} is considered for the target \cite{ennasr2016distributed,ennasr2020time,bar2004estimation,gustafsson2002particle}. 
\begin{align}  \label{eq_targ}
		\mb{x}_{k+1} = F\mb{x}_k+G\mb{q}_k
\end{align}
where $k$ is the time index, $\mb{q}_k=\mc{N}(0,Q)$ is some random input (noise), $\mb{x} = \left(p_x;p_y;p_z;\dot{p}_x;\dot{p}_y;\dot{p}_z\right)$ denotes the target state (position and velocity) in 3D space, $F$ and $G$ represent the transition (system) and input matrices (see details in \cite{gustafsson2002particle}),
\begin{align}  \label{F_ncv}
		F = \left(
		\begin{array}{cc}
			\mb{I}_3 & T \mb{I}_3\\
			\mb{0}_3 & \mb{I}_3 \\ 	
		\end{array} \right),
		G = \left(
		\begin{array}{c}
			\frac{T^2}{2} \mb{I}_3 \\
			T\mb{I}_3 \\ 	
		\end{array} \right)
\end{align}

\subsection{The Sensor Network}
We consider a network of $N$ sensors each located at $\mb{p}_i = (p_{x,i};p_{y,i};p_{z,i})$ (with $i$ as the sensor index) with different $x,y,z$ coordinates. The measurement scenario is as follows. Every sensor receives a beacon signal from the target (with \textit{propagation speed} $c$) and finds its range from $\|\mb{p}(k)-\mb{p}_i(k)\| = ct_i$ with $t_i$ denoting the time-of-arrival (TOA) value (see Fig.~\ref{fig_radar_drone}).  The sensors directly share these measurement values over an undirected network $\mc{G}$ with the set of $\mc{N}_i$ other sensors (referred to as the \textit{neighboring sensors}). The sensors know the neighbors' positions $\mb{p}_j, j\in \mc{N}_i$ as well. The adjacency matrix $W=[w_{ij}]$ of this graph topology $\mc{G}$ is defined as follows; the entry $0<w_{ij}<1$ represents the weight associated with the link $(j,i)$ between sensors $i,j$ and $0$ otherwise. In this work, we assume bidirectional links with symmetric weights, i.e., $W$ is symmetric.  For consensus purposes this symmetric weight-matrix $W$ needs to be stochastic, i.e., $\sum_{i=1}^N w_{ij} = \sum_{i=1}^N w_{ji} = \mb{1}$ \cite{rikos2014distributed,Olfati-2003}. Such weigh-design can be done, for example, via the algorithm in \cite{themis_stochastic}.

\begin{figure} [t]
		\centering
 		\includegraphics[width=2.5in]{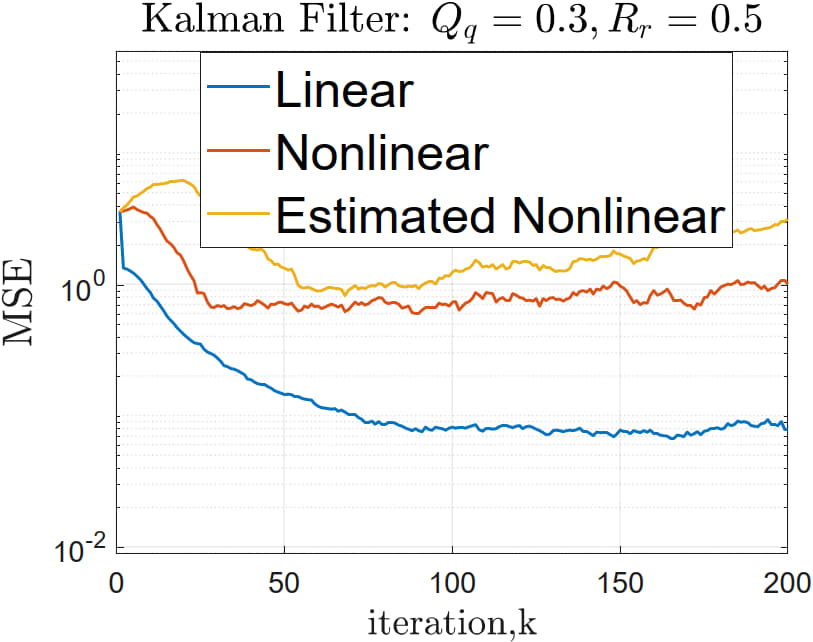}
		\caption{A comparison on the Kalman Filtering performance between our linear measurement model \eqref{eq_h_simp}-\eqref{eq_tdoa_new} and the (linearized) nonlinear model \eqref{eq_tdoa}-\eqref{eq_h_nonlin} \cite{vtc_tracking}. For the latter case, instead of the exact target position $\mb{p}$ (which is unknown), its estimated value $\widehat{\mb{p}}^i$ is needed in \eqref{eq_tdoa}-\eqref{eq_h_nonlin}. This inaccuracy worsens the nonlinear TDOA performance (as compared by the red and yellow curves). }
		\label{fig_kf}
\end{figure}

The nonlinear setup proposed in \cite{ennasr2020time} is $\mb{y}_k^i = h_k^i + \nu_k^i$ (at sensor $i$) with $\nu_k^i=\mc{N}(0,R)$ as general additive Gaussian noise and $h_i(k)$ as the column concatenation of the TDOA measurements $h_{i,j}(k)$ defined as,
\begin{align} 
 \label{eq_tdoa}
		h_{i,j}(k) = \|\mb{p}(k)-\mb{p}_i(k)\|-\|\mb{p}(k)-\mb{p}_{j}(k)\|,~j\in \mc{N}_i.
\end{align}
This is linearized as $\mb{y}^i_k = H_k^i \mb{x}_k + \nu_k^i$ with  measurement matrix $ H_k^i$ of size $\mc{N}_i$-by-$6$ defined  as  the column concatenation of \cite{ennasr2020time},
\begin{align} \nonumber
		h_{i,j}(k) = \Big(
			&\frac{p_x-p_{x,i}}{\|\mb{p}-\mb{p}_i\|}-\frac{p_x-p_{x,j}}{\|\mb{p}-\mb{p}_{j}\|}  , \frac{p_y-p_{y,i}}{\|\mb{p}-\mb{p}_i\|}-\frac{p_y-p_{y,j}}{\|\mb{p}-\mb{p}_{j}\|} , \\ \label{eq_h_nonlin}
			&\frac{p_z-p_{z,i}}{\|\mb{p}-\mb{p}_i\|}-\frac{p_z-p_{z,j}}{\|\mb{p}-\mb{p}_{j}\|},0,0,0\Big),
\end{align}
In our previous works \cite{vtc_tracking,tase}, we proposed a linear measurement model with  
\begin{align} \label{eq_h_simp}
		h_{i,j}(k) =& \frac{1}{2}\Big(\|\mb{p}(k)-\mb{p}_i(k)\|^2-\|\mb{p}(k)-\mb{p}_{j}(k)\|^2\Big)  \\
       =& H^i \mb{p}(k) - \frac{1}{2}(\|\mb{p}_j\|^2 - \|\mb{p}_i\|^2) 
       \label{eq_h_simp2}
	\end{align}	\normalsize
which gives the linear measurement matrix (after some manipulation and simplifications by removing the known bias term) as column concatenation of row vectors in the form
	\begin{align} \label{eq_tdoa_new}
		h_{i,j} = \Big( p_{x,j,i} , p_{y,j,i} ,  p_{z,j,i}, 0 , 0, 0 \Big),
	\end{align}	
with the vector $p_{x,j,i} , p_{y,j,i} ,  p_{z,j,i}$ denoting the $\mb{p}_{j,i}$ as shown in Fig.~\ref{fig_radar_drone}.
This $H^i$ matrix is independent of the target location and is time-invariant. The Kalman Filtering performance of the two scenarios are compared in Fig.~\ref{fig_kf} as an example.

\subsection{The Time-Delay Model} \label{sec_delay}
We assume that the data exchange over every transmission link $(j,i)$ between two sensors $i,j$ is subject to latency. The delay assumed to be bounded by a non-negative integer value $\overline{\tau} \geq \tau_{ij} \geq 0 $. This global max delay $ \overline{\tau} <\infty$ is of finite value to imply no packet drop and loss of information. We assume that the data packets are \textit{time-stamped} so that the recipient knows the time step the data was sent, e.g., via a global discrete-time clock. Our time-delay model and notions mainly follows from \cite{themis_delay,est_delay}.
Over the sensor network $\mc{G}$ of size $N$, define the \textit{augmented state} vector $\underline{\mb{x}}_k = \left(\mb{x}_k; \mb{x}_{k-1}; \dots; \mb{x}_{k-\overline{\tau}} \right)$, and $\mb{x}_{k-r} = \left({x}^1_{k-r};  \dots; {x}^n_{k-r} \right)$ for $0 \leq r \leq \overline{\tau}$. For $N$-by-$N$ adjacency matrix $W$ of the network $\mc{G}$ and known max delay $\overline{\tau}$, similarly define the \textit{augmented matrix} $\overline{W}$ (of size $N(\overline{\tau}+1)$) as
\begin{align} \label{eq_aug_W} 
	\overline{W}= \left( 
	\begin{array}{cccccc}
		W_0 & W_1 & W_{2} & \hdots & W_{\overline{\tau}-1} & W_{\overline{\tau}} \\
		I_N &   0_N & 0_N &\hdots  & 0_N& 0_N\\
		0_N & I_N & 0_N &  \hdots  & 0_N & 0_N  \\
		0_N &  0_N & I_N  &  \hdots  & 0_N & 0_N  \\
		\vdots & \vdots & \vdots & \ddots & \vdots & \vdots \\
		0_N & 0_N & 0_N & \hdots & I_N & 0_N
	\end{array}	
	\right), 
\end{align} 
The block matrices $W_r$ with non-negative entries are defined based on the delay $0\leq r \leq \overline{\tau}$ at different links as
\begin{align}
	W_r(i,j) = \left\{
	\begin{array}{ll}
		w_{ij}, & \text{If}~ \tau_{ij}=r  \\
		0, & \text{Otherwise}.
	\end{array}\right.
\end{align}
Introducing the indicator function $\mb{I}_{k,ij}(r)$  as
\begin{align}
	\mb{I}_{k,ij}(r) = \left\{
	\begin{array}{ll}
		1, & \text{if}~ \tau_{ij}=r  \\
		0, & \text{otherwise}.
	\end{array}\right.
\end{align}
we have $W_r(i,j) = w_{ij} \mb{I}_{k,ij}(r) $ at every time step $k$.
We assume invariant (fixed) delays over time at every link $(j,i)$ but heterogeneous at different links. This assumption (on fixed delays) implies that, at every time $k$, \textit{only one of the entries $W_0(i,j),W_1(i,j), \hdots, W_{\overline{\tau}}(i,j)$ is  equal to  $w_{ij}$   and the other $\overline{\tau}$ terms are zero}. Therefore, the row-sum of the first block ($N$ rows) of $\overline{W}$ and $W$ are equal (and both are row stochastic), i.e., for $1 \leq i \leq N$, $\sum_{j=1}^{N(\overline{\tau}+1)} \overline{w}_{ij} =  \sum_{j=1}^{N} w_{ij} = 1$ and $W = \sum_{r=0}^{\overline{\tau}} {W}_r$ for $k \geq 0$.
Note that, in this work, this large augmented matrix $\overline{W}$ is only introduced to simplify the notations and mathematical analysis and is not needed at any sensor node for filtering purposes. To summarize, 
the followings hold for time-delay $\tau_{ij}$ at every bidirectional link $(j,i)$:
	 \begin{enumerate} [(i)]
	 	\item 
	 	This time-delay  $\tau_{ij} \leq \overline{\tau}$ is known. This upper-bound $\overline{\tau}$ simply means that the sent message from sensor $j$ at time
	 	$k$ eventually reaches  $i$ before or at $k+\overline{\tau}$.  
	 	\item Delay $\tau_{ij}$ is arbitrary, time-invariant, and differs at different links (heterogeneous). Our results also hold for homogeneous (the same) delays at all links. 
	 \end{enumerate}

To justify the above assumptions, suppose  the upper-bound on the delays (or the delay probability) is known, i.e., probability $\mathbb{P}(r)$ for $r\leq\overline{\tau}$ and zero for values above $\overline{\tau}$.
Even though the exact distribution (or the precise delay values) at links $(j,i)$ are unknown (or time-varying), sensors $i,j$ may know $\overline{\tau}_{ij} := \max{\tau_{ij}(k)}$ as the max (possible) delay over their shared communication link $(j,i)$ and both choose to process (the shared information) after $\overline{\tau}_{ij}$ steps (of system dynamics). This ensures that the delayed data-packets at both sides of the link are certainly received by both sensors and they both can update simultaneously. This supports the assumption (ii) on the fixed (time-invariant) delays at every link. This also holds for our assumption on the bounded global max delay $\overline{\tau}$ on all links. See \cite{themis_delay} and \cite[Section~III-D]{est_delay} for details.
For static sensors at fixed positions which communicate, e.g., over a wireless sensor network, it is typical to assume constant  delays (proportional to their distance values), see \cite{AKYILDIZ2002393,liu2019survey} for details.

\section{The Main Results} \label{sec_main}
Every sensor $i$ updates its estimation as follows; it performs one iteration of consensus processing on the filtering information received from sensors $j\in \mc{N}_i$ as they arrive. Note that these information are subject to \textit{known time-delays} (as discussed in Section~\ref{sec_delay}), e.g., when the data packet is time-stamped. Second, sensor $i$ updates this \textit{a-priori} estimate $\widehat{\mb{x}}^i_{k|k-1}$ by its TDOA measurement $y^k_i$ (referred to as the \textit{innovation-update}) which gives the \textit{posterior} estimate $\widehat{\mb{x}}^i_{k|k}$. The distributed (and local) filtering protocol for tracking is formulated as, 

\small \begin{align}\label{eq_p} 
	\widehat{\mb{x}}^i_{k|k-1} =& w_{ii}F\widehat{\mb{x}}^i_{k-1|k-1} + \sum_{j\in\mathcal{N}_i} \sum_{r=0}^{\overline{\tau}} w_{ij}F^{r+1}\widehat{\mb{x}}^j_{k-r|k-r} \mb{I}_{k-r,ij}(r),
	\\\label{eq_m}
	\widehat{\mb{x}}^i_{k|k} =& \widehat{\mb{x}}^i_{k|k-1} + K_i H^{i\top} \left(\mb{y}^i_k-H^i\widehat{\mb{x}}^i_{k|k-1}\right),
\end{align} \normalsize 
This filtering protocol, as mentioned in Section~\ref{sec_intro}, is in single time-scale, i.e., between every $k-1$ and $k$ steps of the target dynamics only one step of data-sharing/consensus-fusion occurs. Over every link $(j,i)$, $\mb{I}_{k-r,ij}(r)$ is non-zero \textit{only} for one $r \in [0~\overline{\tau}]$ which follows from fur fixed delay assumption. 

In compact form, define $\widehat{\mb{x}}_{k|k-1} :=(\widehat{\mb{x}}^1_{k|k-1}; \dots; \widehat{\mb{x}}^N_{k|k-1})$ which denotes the column-concatenation of $\widehat{\mb{x}}^i_{k|k-1}$s and the augmented state estimate vectors are
$\underline{\widehat{\mb{x}}}_{k|k-1} = \left(\widehat{\mb{x}}_{k|k-1}; \widehat{\mb{x}}_{k-1|k-2}; \dots; \widehat{\mb{x}}_{k-\overline{\tau}|k-\overline{\tau}-1} \right)$ and similarly for  $\widehat{\mb{x}}_{k|k}$ and $\underline{\widehat{\mb{x}}}_{k|k}$. To simplify the formulation, the augmented version of \eqref{eq_p}-\eqref{eq_m} is given as,

\small \begin{align}\label{eq_p_aug}
	\underline{\widehat{\mb{x}}}_{k|k-1} =&\overline{WF} \underline{\widehat{\mb{x}}}_{k-1|k-1},
	\\\label{eq_m_aug}
	\underline{\widehat{\mb{x}}}_{k|k} =&  \underline{\widehat{\mb{x}}}_{k|k-1} + \mb{b}^{\overline{\tau}+1}_1 \otimes K D_H^\top \left(\mb{y}_k-D_H\Xi^{Nn}_{1,\overline{\tau}}\underline{\widehat{\mb{x}}}_{k|k-1}\right),
\end{align} \normalsize
with
${D}_H:=\mbox{diag}[H^i]$, $K := \mbox{diag}[K_i]$, and the (auxiliary) matrix 
$\Xi^m_{i,\overline{\tau}}= (\mb{b}^{\overline{\tau}+1}_i \otimes I_m)^\top $
with $\mb{b}^{\overline{\tau}+1}_i$ as the unit (column) vector of $i$th (Cartesian) coordinate, $1 \leq i \leq {\overline{\tau}+1}$. 
The matrix $\overline{WF}$ represents \textit{modified} augmented version of $W \otimes F$ as,

\scriptsize \begin{align} \label{eq_aug_WA}
	\overline{WF} := \left( 
	\begin{array}{cccccc}
		W_0 \otimes F & W_1\otimes F^2  &  \hdots & W_{\overline{\tau}-1}\otimes F^{\overline{\tau}}  & W_{\overline{\tau}} \otimes F^{\overline{\tau}+1}  \\
		I_{Nn} &   0_{Nn}  &\hdots  & 0_{Nn}& 0_{Nn}\\
		0_{Nn} & I_{Nn} &   \hdots  & 0_{Nn} & 0_{Nn}  \\
		0_{Nn} &  0_{Nn} &  \ddots  & 0_{Nn} & 0_{Nn}  \\
		\vdots & \vdots &  \hdots & \vdots & \vdots \\
		0_{Nn} & 0_{Nn} &  \hdots & I_{Nn} & 0_{Nn}
	\end{array}	
	\right),
\end{align} \normalsize
Similarly, define the (augmented) error vector as $\underline{\mb{e}}_{k}  := \underline{\mb{x}}_{k} - \underline{\widehat{\mb{x}}}_{k|k}$ with the  $\underline{\mb{x}}_{k} := \left(1_{N} \otimes   \mb{x}_k; 1_{N} \otimes \mb{x}_{k-1}; \dots; 1_{N} \otimes \mb{x}_{k-\overline{\tau}} \right) $  as the augmented state. 
The error dynamics then follows as,
\begin{align} 
	\underline{\mb{e}}_{k} =  \overline{WF}
- \mb{b}^{\overline{\tau}+1}_1 \otimes K \overline{D}_H \Xi^{Nn}_{1,\overline{\tau}}\overline{WF} + \underline{\mb{\eta}}_k 
=: \underline{\widehat{F}} \underline{\mb{e}}_{k-1}  + \underline{\mb{\eta}}_k, \label{eq_err1}
\end{align}
with $\underline{\widehat{F}}$ denoting the \textit{closed-loop matrix} associated with the global augmented observer error dynamics, $\overline{D}_H= D_H^\top D_H$, and $\underline{\mb{\eta}}_k$ collecting the noise terms, see more details in \cite{est_delay}.
Putting $\overline{\tau}=0$, gives $\widehat{F}=W\otimes F - K \overline{D}_H (W\otimes F)$ as the \textit{delay-free closed-loop matrix} \cite{jstsp,jstsp14}. In our previous work \cite{est_delay,jstsp14} we derived the condition for Schur stability of the error dynamics \eqref{eq_err1}, i.e., to have $\rho(\underline{\widehat{F}})<1$ (and $\rho(\widehat{F})<1$). First, recall from \cite{est_delay,jstsp,usman_cdc:11} that for stability of \eqref{eq_err1} we only need \textit{distributed observability}, i.e., the pair $(W\otimes F, \overline{D}_H)$ to be observable (or detectable). This implies observability over network $\mc{G}$ (with adjacency matrix $W$), and is easier to satisfy rather than the local observability \cite{mohammadi2015distributed}.  
In addition, from the definition of $\overline{WF}$, its row-stochasticity holds for any row stochastic $\overline{W}$ matrix. This ensures that the consensus term in \eqref{eq_p_aug} leads to proper averaging of the (possibly delayed) filtering data over the network. 

In \cite{jstsp,jstsp14}, we proved that given a full-rank system matrix $F$ with observable output matrix $H$, one can ensure distributed observability over any strongly-connected (SC) network. In other words, given $\mbox{S-rank}(F)=\mbox{dim}(F)$, $(F,H)$-observability  (this is global not local observability), and $W$ to be a bi-stochastic irreducible adjacency matrix, then $(W\otimes F, \overline{D}_H)$-observability is guaranteed (in structural or generic sense\footnote{Note that the proofs and results are based on structured systems theory (also known as generic analysis) \cite{woude:03,liu-pnas}. This implies that our observability results hold for almost all arbitrary entries of independent system, measurement, and consensus matrices as far as their structure (fixed zero-nonzero pattern) is unchanged.}).
Note that strong-connectivity intuitively implies that the estimation data of each sensor eventually reaches every other sensor via a \textit{connected sequence of sensor nodes}. This is an easier condition to satisfy instead of data sharing in the direct neighborhood of sensors.

For an observable pair $(W\otimes F, \overline{D}_H) $, we showed in \cite[Section~III]{est_delay} that the same LMI-based gain matrix $K$ designed for the delay-free case ($\overline{\tau}=0$) to ensure $\rho(\widehat{F})<1$, may also ensure Schur stability of $\rho(\underline{\widehat{F}})<1$ for the error dynamics \eqref{eq_err1} under certain bounds on the time-delay.  
This is summarized in the following theorem.  

\begin{thm} \label{thm_tau}
	Given $(W\otimes F, \overline{D}_H) $-observability and  proper feedback gain matrix $K$ such that $\rho({\widehat{F}})<1$, then $\rho(\underline{\widehat{F}})<1$ holds for any   $\overline{\tau} \leq \overline{\tau}^*$, with $\overline{\tau}^*$  satisfying, 
	\begin{align} \label{eq_tau*}
	    \rho(W\otimes F^{\overline{\tau}^*+1} - K \overline{D}_H (W\otimes F^{\overline{\tau}^*+1}) )< 1.
	\end{align}  
\end{thm}

In fact, this theorem provides a sufficient bound on $\overline{\tau}$ to ensure stable tracking (in the presence of maximum possible delays $\tau_{ij} = \overline{\tau}$). 
For any maximum delay $\overline{\tau}$ satisfying Eq.~\eqref{eq_tau*}, the tracking error under filtering protocol \eqref{eq_p}-\eqref{eq_m} is guaranteed to decay over time and remains bounded steady-state. This is better illustrated by simulations in Section~\ref{sec_simi}. Note that the above theorem is stated for general (possibly) \textit{unstable} systems with $\rho(F)>1$. For the given NCV dynamics we have  $\rho(F)=1$ which implies that the above theorem is easier to satisfy as Eq.~\eqref{eq_tau*} may hold for larger values of $\overline{\tau}^*$. In fact, one can show that, since $\rho(W)=1$ and $\rho(F^{\overline{\tau}^*}) = \rho(F)=1$, the proposed filtering algorithm is (almost) delay-tolerant and the error stability is guaranteed for even large values of $\overline{\tau}$ (via proper feedback gain design as in \cite[Section~III-A]{est_delay}).

Note that the LMI design of the \textit{block-diagonal} gain matrix $K$ follows from both structure and numerical values of $F$ and $H^i$ matrices. Recall that, the (linearized) nonlinear model  \eqref{eq_tdoa}-\eqref{eq_h_nonlin} gives a time-varying $H^i$ (since it is a function of $\mb{p}$ as the target position). This time-dependency is another drawback of the nonlinear model \cite{ennasr2020time} as it needs repetitive LMI-design of $K$ at every iteration $k$. This is computationally burdensome and less real-time feasible. In contrast, our proposed linear model \eqref{eq_h_simp}-\eqref{eq_tdoa_new}, gives a time-invariant measurement matrix $H^i$, and thus, the LMI $K$-design is done only once at the initialization of the algorithm. This significantly reduces the computational load on the sensors. Further, the $K$-design is less accurate in the nonlinear measurement case, because it is based on the \textit{estimated position} of the target  $\widehat{\mb{p}}_i$. This further increases the inaccuracy in the gain and the tracking error, see more illustrations in \cite[Section~V]{vtc_tracking}. 

In addition, our proposed distributed and local filtering method allows for simultaneous fault-detection and isolation (FDI) schemes as \cite[Algorithm~1]{vtc_tracking}. See also \cite{Ugrinovskii8742899,deghat2019detection} for distributed estimation resilient to biasing attacks.
Some more example scenarios for such \textit{fault-tolerant design} (both stateful and stateless) and \textit{survivable network design} in distributed observer setup are further discussed in \cite{anomaly_ECC22}. These algorithms enable each sensor to locally detect and isolate its biased (or faulty) TDOA measurement and prevent the cascade of faulty data among all sensors.

\section{Simulations} \label{sec_simi}
\begin{figure*} [t]
	\centering
	\includegraphics[width=2.15in]{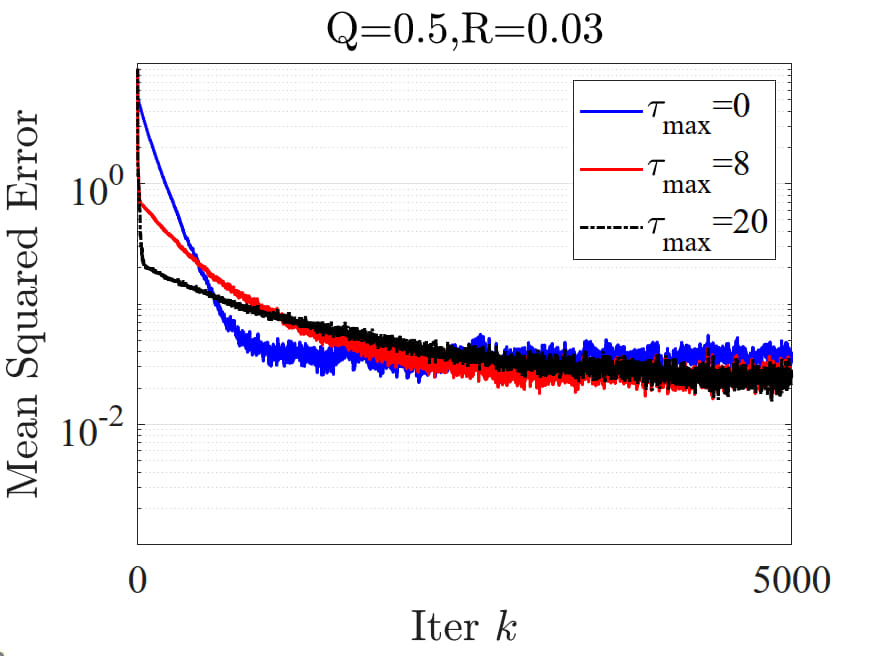}
	\includegraphics[width=2.15in]{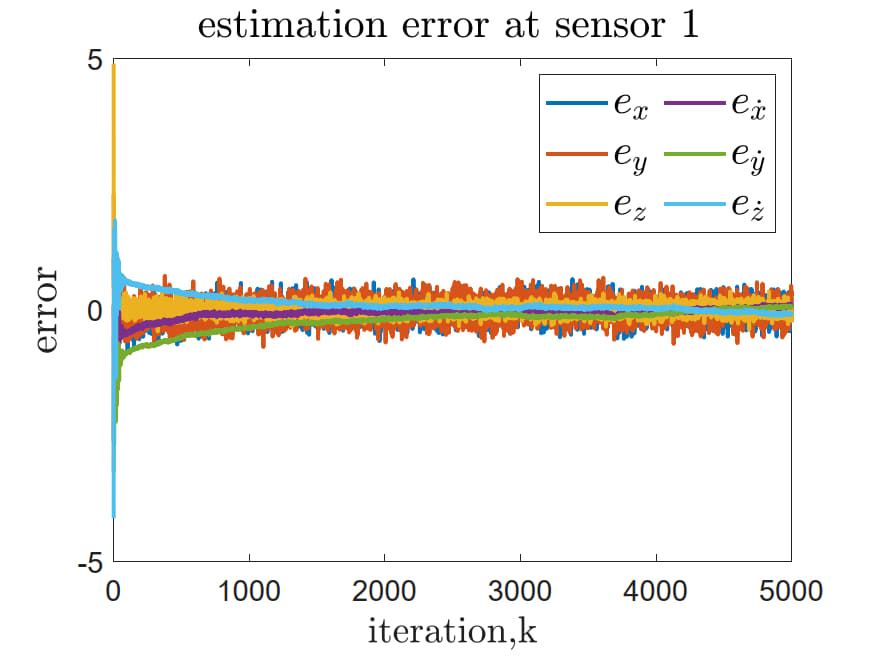}
	\includegraphics[width=2.2in]{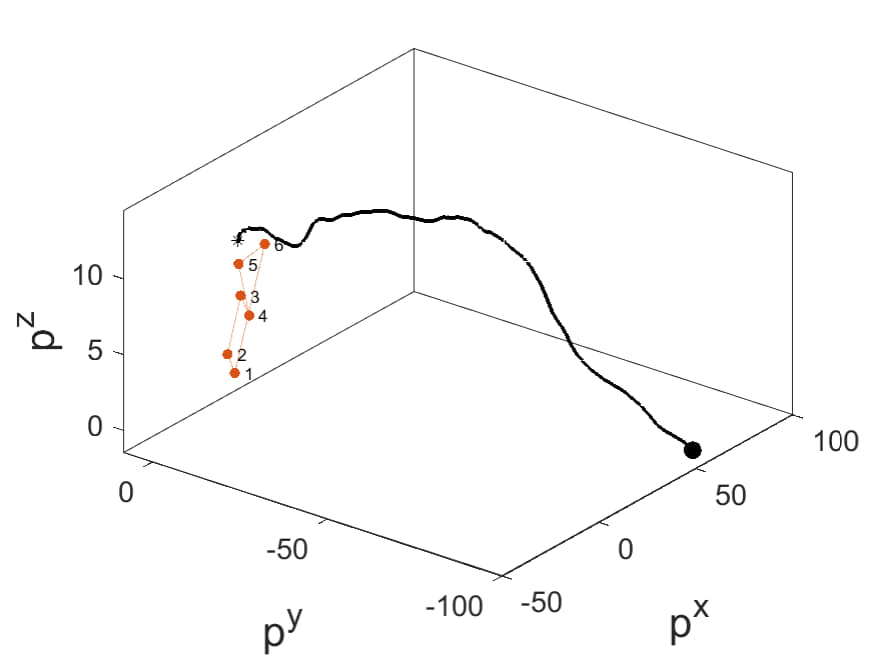}
	\includegraphics[width=2.15in]{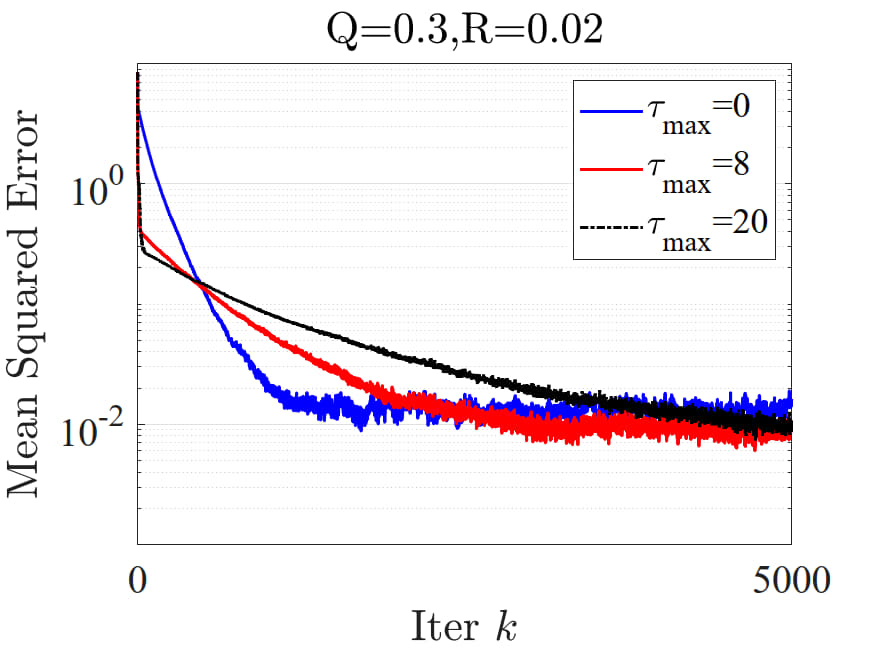}
	\includegraphics[width=2.15in]{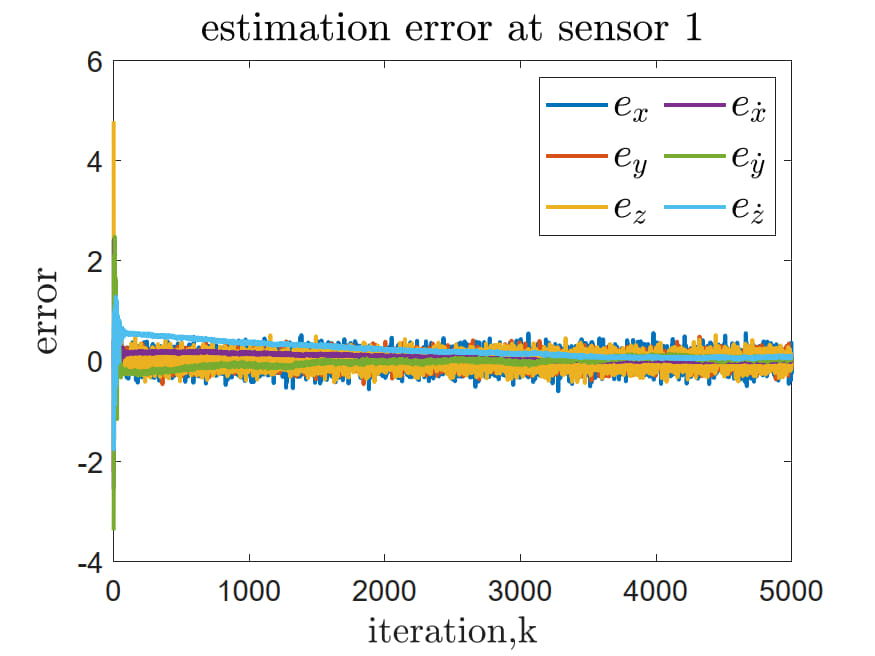}
	\includegraphics[width=2.2in]{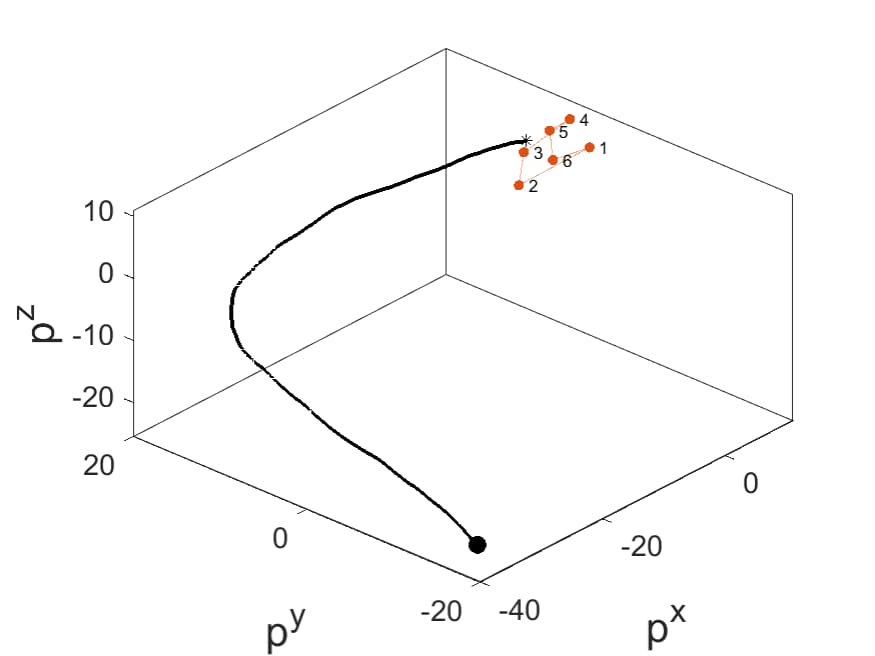}	
	\includegraphics[width=2.15in]{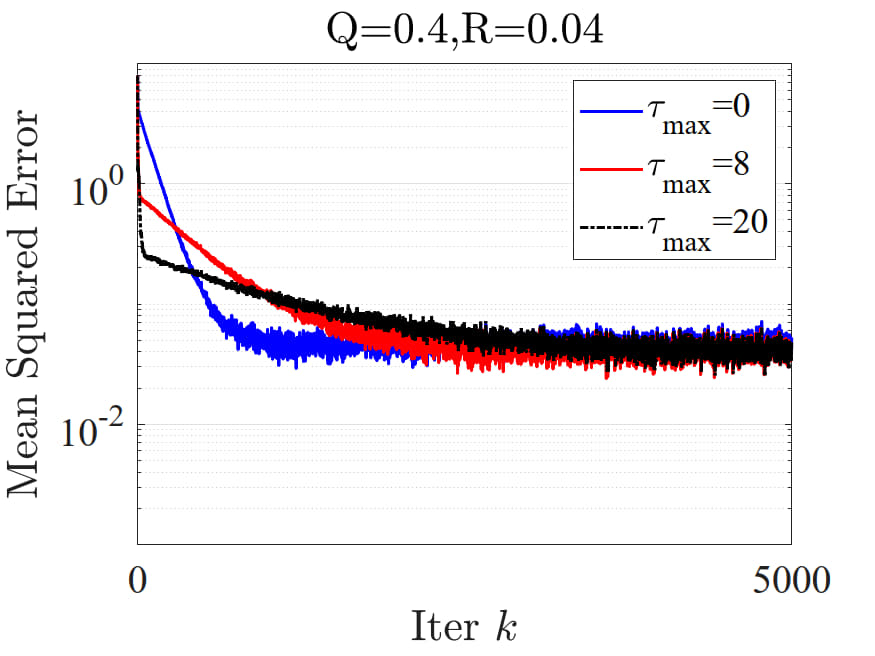}
	\includegraphics[width=2.15in]{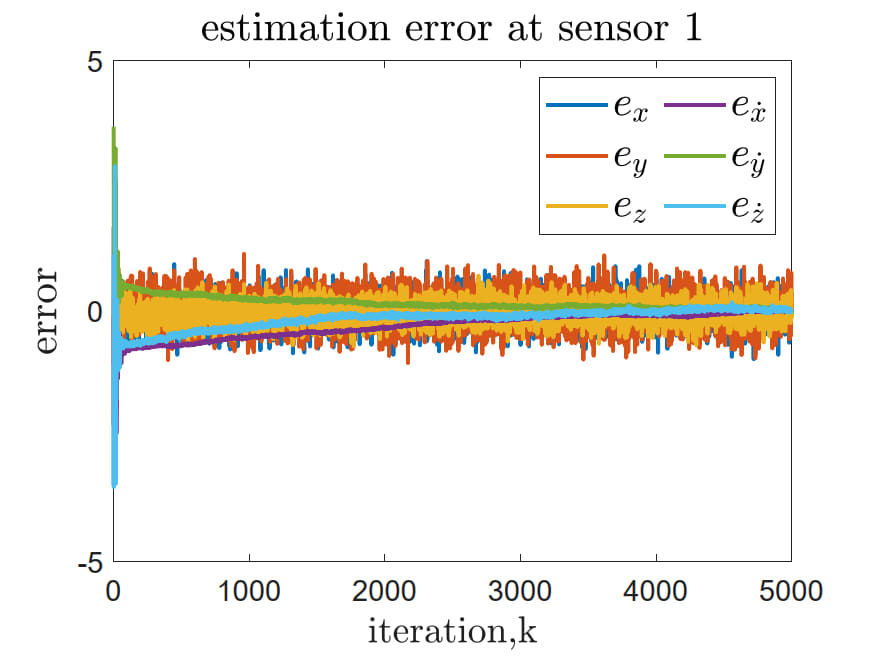}
	\includegraphics[width=2.2in]{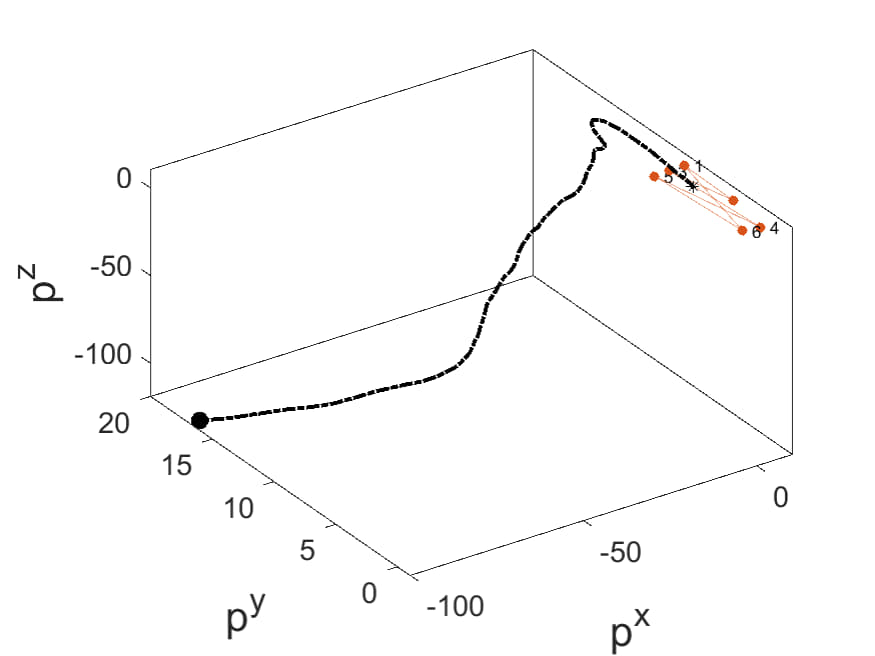} 		
	\caption{The MC performance of the proposed distributed filtering \eqref{eq_p}-\eqref{eq_m} over a cyclic network of $N=6$ static sensors (shown by orange in the right figure). Different upper-bound $\overline{\tau}=0,8,20$ on the time-delays are considered, where $\overline{\tau}=0$ implies the ideal case with no latency. The evolution of state errors at sensor $1$ are also shown for $\overline{\tau} = 20$. The target path in 3D space is shown by black (solid line) and the large black circle shows its final position after certain time iterations.}
	\label{fig_df} 
\end{figure*}
For the simulations a cyclic network of $N=6$ sensors (which is SC) tracks a randomly maneuvering single target with NCV dynamics \eqref{eq_targ}-\eqref{F_ncv}. The TDOA measurements follows from the linear model \eqref{eq_h_simp}-\eqref{eq_tdoa_new}. Clearly, since the $H_i$ matrices are time-invariant and independent of the target state $\mb{p}$, the distributed filtering can be applied more conveniently and in a more accurate way (as discussed in details in the previous sections).  The consensus fusion weights are chosen arbitrarily such that to satisfy the symmetric and stochastic properties \cite{themis_stochastic}.
We choose random initial locations (in the range $[0~10]$) for the sensors and the target. Then, the distributed filter \eqref{eq_p}-\eqref{eq_m} is applied  over this setup for $10$ Monte-Carlo (MC) trials. The MC mean square estimation error (averaged at all sensors and states) is shown in Fig.~\ref{fig_df}, which is unbiased in steady-state. We chose different maximum time-delay $\overline{\tau} = 0,8,20$ and different Gaussian noise statistics with variances $Q$ and $R$ given in the figure titles. Due to space limitations, as an example, the estimation errors (of all $6$ position and velocity states) are only given for one of the $6$ sensors. For this setup, we used the LMI $K$-design in \cite{est_delay} that gives a separate $6 \times 6$ feedback gain matrix $K_i$ for each of the $6$ sensors (which are not given here due to space limitations). For this gains  $\rho(\underline{\widehat{F}}) = 0.99<1$, and the results of Theorem~\ref{thm_tau} holds. The errors, as illustrated in Fig.~\ref{fig_df}, are bounded steady-state stable implying that the protocol works under different max delay values $\overline{\tau}$. The target path and the sensors' positions are also shown in Fig. \ref{fig_df}(Right). The remaining simulation parameters are given in the figure caption.  

\section{Conclusions and Future Works} \label{sec_conc}
This paper provides a \textit{linear} TDOA measurement model for distributed estimation that improves accuracy of the  existing nonlinear TDOA models. The results can be further extended to incorporate \textit{local} stateless or stateful FDI strategies to detect and isolate (possible) faults or anomalies locally. 
Considering mobile sensors (AR drones or UAVs) in our linear TDOA setup is more challenging as it requires certain rigid formation strategies to fix the distances between the sensors \cite{tase}. This is to ensure the time-independence of the measurement matrix $H^i$ and also to keep the maneuvering target in the detection range of the sensors (or agents). 
The notions of cost-optimal and energy-efficient design as other future research directions \cite{pequito2014optimal,spl18}.




\bibliographystyle{IEEEbib}
\bibliography{bibliography}

\end{document}